\documentstyle[preprint,aps,psfig]{revtex}
\begin{document}

\draft
\tightenlines
\preprint{\vbox{\hbox{U. of Iowa 99-2502}}}
\title{Non-Perturbative Fine-Tuning in Approximately Supersymmetric Models}
\author{Y. Meurice\\ 
{Dept. of Physics and Astr., Univ. of Iowa, Iowa City, Iowa 52242, USA}}
\maketitle

\begin{abstract}
We present two Fermi-Bose models with an approximate supersymmetry
and which can be solved numerically 
with great accuracy using the renormalization group method.
The bosonic parts of these models consist in Dyson's hierarchical model
with one and two scalar components respectively.
We discuss the question of the perturbative 
cancellations of divergences and compare with 
the non-perturbative fine-tunings necessary to keep the 
renormalized scalar mass small in cut-off units. 
We show evidence for non-perturbative cancellations of quantum corrections,
however, we were not able to achieve exact cancellations without fine-tuning.

\end{abstract}

\narrowtext
\newpage

The fact that the bare parameters of a scalar field theory require a
fine-tuning in order to keep the renormalized mass small in cut-off units
is usually regarded as an argument against fundamental scalars\cite{susskind}.
A possible resolution of this inelegant feature consists in adding 
degrees of freedom in such a way that the
quantum fluctuations cancel, making small scalar masses a more 
natural outcome. 

From a non-perturbative renormalization group\cite{wilson} analysis, 
a relevant direction
is necessary in order to describe massive particles, 
making the fine-tuning process
unavoidable. This situation can be seen very simply in the 
case of a free scalar theory with a cut-off. In this example,
the fact that the bare mass ($m_B$) has to be small (in cut-off units)
in order to get a small physical mass ($m_R$) is obvious since
these quantities are identical. 
On the other hand in an interacting scalar theory,
we usually need to take $m_B^2$ negative and large in absolute value 
and also to
adjust many digits of this quantity in order to get a small $m_R$.
The difference between the two situations is that in the second case,
there is no bare quantity which controls the size of $m_R$.
One would like to understand under which circumstances
the inclusion of fermions allows us to obtain a small $m_R$ whenever
we choose a small $m_B^2$. 

There are known four-dimensional examples\cite{wz} where one can 
cancel the perturbative
quadratic divergences by imposing simple relations between 
the Yukawa couplings and the scalar quartic couplings. 
However, it is not clear that there exists a non-perturbative
regularization which fully preserves the perturbative naturalness.

In the following, we present two models where there is an equal number
of fermions and bosons and which can be solved
non-perturbatively with great accuracy\cite{finite}. 
The bosonic part of these model
is a Dyson's hierarchical model\cite{dyson}. 
In this model, the renormalization group
transformation maps the local measure into another local measure.
The price to pay for this simplifying feature is that the kinetic term
is not ultra-local. The free action for $N$ massless scalar fields 
$\phi_x^{(i)}$ reads
\begin{equation}
S_B^{free}={1\over 2}\sum_{x,y,i}\phi_x^{(i)}D^2_{xy}\phi_y^{(i)}\ ,
\end{equation}
where $x$ and $y$ run over the sites and $i$ from 1 to $N$. 
The explicit form of $D^2_{xy}$ is given below in Eq. (\ref{eq:ham}).
The action for free massless fermions reads
\begin{equation}
S_F^{free}=\sum_{x,y,i}\bar{\psi}_x^{(i)}D_{xy}\psi_y^{(i)}\ ,
\end{equation}
where the $\psi_x^{(i)}$ and  $\bar{\psi}_x^{(i)}$ are Grassmann numbers 
integrated with a measure 
\begin{equation}
\int\prod_{x,i}d\psi_x^{(i)}d\bar{\psi}_x^{(i)} \ .
\end{equation}
As indicated by the notation, we have
\begin{equation}
D_{xy}^2=\sum_zD_{xz}D_{zy} \ .
\label{eq:sq}
\end{equation}

The free action $S_B^{free}+S_F^{free}$ is invariant at first order 
under the transformation
\begin{eqnarray}
\nonumber
\delta\phi^{(i)}_x=\epsilon\bar{\psi}_x^{(i)}+{\psi_x^{(i)}}\bar{\epsilon}\\ 
\delta\psi_x^{(i)}=\epsilon\sum_xD_{xy}\phi_y^{(i)}\\
\nonumber
\delta\bar{\psi}_x^{(i)}=\bar{\epsilon}\sum_xD_{xy}\phi_y^{(i)} \ . \\ 
\nonumber
\end{eqnarray}
The $\epsilon$ and $\bar{\epsilon}$ are Grassmann numbers.
Integration by part or Leibnitz's rule cannot be used for $D_{xy}$ and the 
order $\epsilon\bar{\epsilon}$ variations do not cancel.

We now give the explicit form of $D^2_{xy}$ at finite volume.
For a  hierarchical model with $2^{n_{max}}$ sites,
we label the sites with $n_{max}$
indices $x_{n_{max}}, ... , x_1$, each index being 0 or 1. In order to
understand this notation, one can divide the $2^{n_{max}}$ sites into
two blocks, each containing $2^{n_{max}-1}$ sites. If $x_{n_{max}}=0$,
the site is in the first box, if $x_{n_{max}} = 1$, the site is in the
second box. Repeating this procedure $n$ times 
(for the two boxes, their respective two sub-boxes, etc... ), 
we obtain an unambiguous
labeling for each of the sites. With these notations,
\begin{equation}
S_B^{free} =
-{\beta_B}\sum_{n=1}^{n_{max}}({c_B\over4})^n\sum_{x_{n_{max}},...,x_{n+1},i} 
(\sum_{x_n,...,x_1}\phi^{(i)}_{(x_{n_{max}},...x_1)})^2
+{\beta_B c_B\over{2-c_B}}\sum_{x_{n_{max}},...,x_{n+1},i}( \phi^{(i)}_{(x_{n_{max}},...x_1)})^2\ .
\label{eq:ham}
\end{equation}
The index $n$
corresponds to the interaction of the total field in blocks of size $2^n$.
The constant $c_B=2^{1-2/D}$ is a free parameter which controls the decay 
of the 
iterations with the size of the boxes and 
can be adjusted in order to mimic a $D$-dimensional model. 
Similarly the free massless fermionic action reads
\begin{eqnarray}
S_F^{free} =
-{\beta_F}\sum_{n=1}^{n_{max}}({c_F\over4})^n\sum_{x_{n_{max}},...,x_{n+1},i} 
(\sum_{x_n,...,x_1}\bar{\psi}^{(i)}_{(x_{n_{max}},...x_1)})
(\sum_{x_n,...,x_1}\psi^{(i)}_{(x_{n_{max}},...x_1)})
\\
+{\beta_Fc_F\over{2-c_F}}\sum_{x_{n_{max}},...,x_{n+1},i}
\bar{\psi}^{(i)}_{(x_{n_{max}},...x_1)}\psi^{(i)}_{(x_{n_{max}},...x_1)}\ ,
\label{eq:hamf}
\end{eqnarray}
with $c_F=2^{1-1/D}. $
Using the techniques explained in \cite{jmp}, one can show that the fermionic
operator is the square root of the bosonic operator (see Eq. (\ref{eq:sq}))
provided that 
\begin{equation}
{\beta_Fc_F\over{2-c_F}}=( {\beta_B c_B\over{2-c_B}})^{1\over 2}
\end{equation}

We now introduce local interactions.
The Grassmann nature of the fermionic fields restricts severely the type
of interactions allowed. For instance, for one flavor ($N=1$), the most
general local measure is 
\begin{equation}
{\cal{W}}(\phi,\psi,\bar{\psi})=W(\phi)+\psi\bar{\psi}A(\phi)
\end{equation}
For convenience, we will always reabsorb the second term of 
Eqs. (\ref{eq:ham})
and (\ref{eq:hamf}) which are local, in the local measure.
In the following calculations, 
$W(\phi)$ will take 
the Landau-Ginzburg (LG) form:
\begin{equation}
W(\phi)\propto e^{-(( {\beta_B c_B\over{2-c_B}})+{1\over 2}m_B^2 \phi^2+ 
\lambda_B\phi^{4})} \ .
\label{eq:lg}
\end{equation}
If the two functions $W$ and $A$ are proportional, the fermionic degrees
of freedom decouple.
The renormalization group transformation takes the form
\begin{eqnarray}
W&\rightarrow & 2 A\star W \\
A&\rightarrow &2\beta_F A\star W +({4\over{c_F}})W\star W
\end{eqnarray}
where the $\star$ operation  means a convolution, a multiplication
by an exponential and a rescaling of the new field. More precisely
\begin{equation}
A\star B(\phi)\equiv  e^{{\beta_B \over 2} ( \phi ^2)}
\int d\phi ' A({{(\phi2c_B^{-{1\over 2}} -\phi ')}\over 2})
B({{(\phi2c_B^{-{1\over 2}} +\phi ')}\over 2}) \ ,
\label{eq:bsp}
\end{equation}
The introduction of a Yukawa coupling can be achieved by allowing a linear term
in $A(\phi)$. Such a term breaks explicitly the $Z_2$ symmetry of the 
LG measure. Such a model is characterized by a sudden change from the symmetric
phase behavior to the broken phase behavior followed by unexpectedly
long low-temperature ``shoulders'' (see \cite{long} for an explanation of 
this terminology).

A richer behavior is observed in the case of the two flavors ($i=1,2$) models.
In the following we have restricted our investigation to the 
type of bilinear coupling appearing in the Wess-Zumino\cite{wz} model, namely
\begin{eqnarray}
\nonumber
{\cal{W}}(\phi^{(i)},\psi^{(i)},\bar{\psi}^{(i)})=
W(\phi^{(i)})
+A(\phi^{(i)})(\bar{\psi}^{(1)}\psi^{(1)}+\bar{\psi}^{(2)}\psi^{(2)})+\\
B(\phi^{(i)})\psi^{(1)}\psi^{(2)}-B^{\star}(\phi^{(i)})
\bar{\psi}^{(1)}\bar{\psi}^{(2)}+
T(\phi^{(i)})\bar{\psi}^{(1)}\psi^{(1)}\bar{\psi}^{(2)}\psi^{(2)} \ .
\end{eqnarray}
For convenience, we again absorb the local parts of Eqs. (\ref{eq:ham}) and 
(\ref{eq:hamf}). 
This is not the most general measure, however it closes under 
the renormalization group transformation which takes the form
\begin{eqnarray}\nonumber
W&\rightarrow & (W\star T +A\star A +B\star B^{\star} )\equiv W'\\ 
\nonumber
A&\rightarrow &\beta_F W' +{4\over{ c _F}}A\star T\\ 
B&\rightarrow &{4\over{ c _F}}B\star T\\ 
\nonumber
T&\rightarrow & {8\over{ c _F^2}}T\star T+ \beta_F {8\over{ c _F}}A\star T+
(\beta_F)^2 W' \ .\\
\nonumber
\end{eqnarray}
In addition we will impose that the function $B$ have the following form:
\begin{equation}
B(\phi^{(i)})=(\phi^{(1)}+i\phi^{(2)})P((\phi^{(1)})^2+(\phi^{(2)})^2) \ ,
\label{eq:bdef}
\end{equation}
while $W$, $A$ and $T$ are $O(2)$-invariant.
The model is then invariant under the R-symmetry
\begin{eqnarray}
\nonumber
(\phi^{(1)}+i\phi^{(2)})&\rightarrow & e^{i\theta}(\phi^{(1)}+i\phi^{(2)})\\
\psi^{(j)}&\rightarrow &e^{-i{\theta\over 2}}\psi^{(j)}\\ 
\nonumber
\bar{\psi}^{(j)}&\rightarrow &e^{i{\theta\over 2}}\bar{\psi}^{(j)} \ .
\end{eqnarray}

We now present three numerical calculations performed with the second model.
In all cases we will set $D=4$ in $c_B$ and $c_F$.
In the following, we have chosen the value of $\beta_B$ and $\beta_F$ in 
such a way that 
\begin{equation}
{\beta_Fc_F\over{2-c_F}}=( {\beta_B c_B\over{2-c_B}})^{1\over 2}=1 \ ,
\end{equation}
in order to make the perturbative expansion more similar to usual
Feynman diagram's calculations.

First, we consider the case where the fermions decouple from the bosons.
$W$ takes a LG form
\begin{equation}
W(\phi)\propto e^{-((( {\beta_B c_B\over{2-c_B}})+{1\over 2}m_B^2 )
\sum_i(\phi^{(i)})^2
+ 
\lambda_B(\sum_i(\phi^{(i)})^2)^2) }\ .
\end{equation}
The value of $m_R^2$, defined as the 
inverse of the zero-momentum two-point function, is shown in Fig. 1. as 
a function of $m_B^2$. These quantities are expressed in 
cut-off units. For reference we have also displayed the one-loop
perturbative result and the trivial gaussian result. One sees that the 
scalar self-interaction moves $m_R^2$ up and $m_R^2\simeq 0.2$ when 
$m_B^2$ goes to zero. The one-loop result is quite good when
$m_R^2$ is large enough but deteriorates when this quantity becomes smaller.

In the second calculation, we consider a bosonic model with 
a bare mass $m_B$ and $\lambda_B=0$ 
coupled to a fermion with the following couplings:
\begin{eqnarray}
\nonumber
A&=&(-1-m_B)W\\
P&=&g_yW\\
\nonumber
T&=&((-1-m_B)^2 + g_y^2((\phi^{(1)})^2+(\phi^{(2)})^2))W \ .
\nonumber
\end{eqnarray}
The results are shown in Fig. 2 for $g_y=\sqrt{0.08}\simeq0.28$. 
One sees that the Yukawa coupling moves $m_R^2$ down.
For $m_B^2\simeq 0.094$, $m_R$ becomes 0 and for smaller of $m_B^2$, we enter
the broken symmetry phase.

We have then repeated the second calculation 
with $\lambda_B=0.01$ 
instead of 
0. 
In perturbation theory, the one-loop quadratic divergence cancel when
$m_B=0$ and 
\begin{equation}
8\lambda_B=g_y^2 \ ,
\end{equation}
which justifies our choice of coupling constant.
The results are shown in Fig. 3.
One sees that the Yukawa coupling in part cancels the effects of the scalar 
self-interaction, however, the cancellation is not as good as in the one-loop
formula where $m_R$ goes to zero when $m_B^2$ goes to zero. Instead, we 
found numerically that $m_R^2\simeq 0.044$ when $m_B^2$ goes to zero.
A summary of the three numerical results is shown in Fig. 4. 

It is possible to fine-tune $g_y$ in order to get $m_R=0$.
An example is shown in Fig. 5 for $\lambda_B=0.01$ and $m_B^2=0.01$.
We see that there exist a critical value of $g_y$ which is approximately 
0.46 and 
where $m_R$ becomes 0. For larger values of $g_y$, we enter the symmetry 
broken phase.  An essentially similar figure is obtained for $m_B^2=0$.
In both cases, the exact critical value of $g_y$ is about 50 percent larger
than the perturbative one.

In conclusion, we have shown that the idea of canceling the 
quantum correction
inspired by perturbation theory have 
qualitatively a non-perturbative counterpart. However, we have not found a way
to make this cancellation very accurate or exact without fine-tuning.

We thank the Institut de Physique Theorique of Louvain-la-Neuve, the 
CERN theory division and the Aspen Center for Physics where part of this
work was completed and B. Oktay for valuable conversations.
This research was supported in part by the Department of Energy
under Contract No. FG02-91ER40664.

\begin{figure}
\vskip50pt
\centerline{\psfig{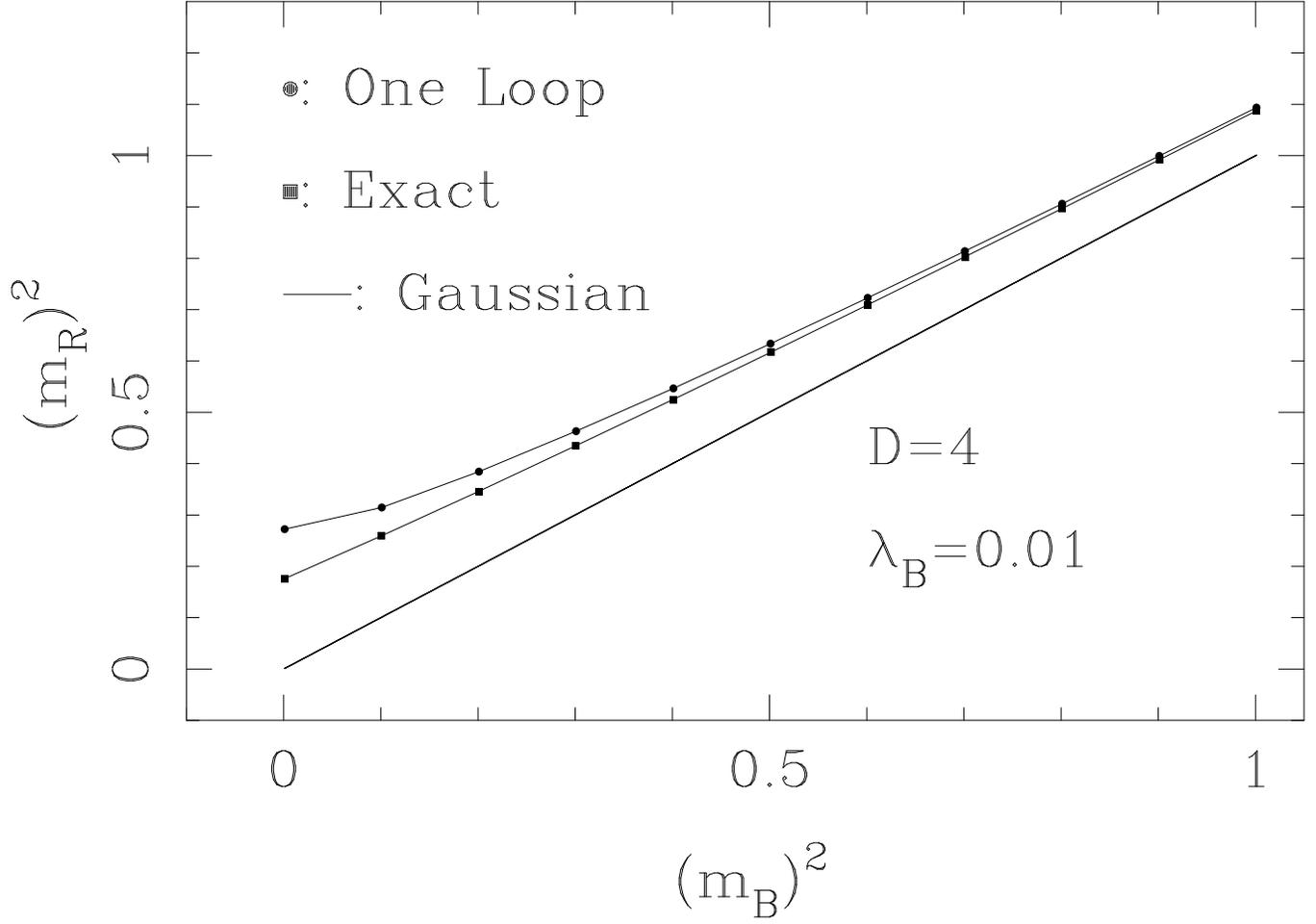}}
\vskip50pt
\caption{
The renormalized mass as a function of the bare mass in a bosonic $O(2)$
model with bare quartic coupling fixed to 0.01.
}
\end{figure}
\vfill
\eject
\begin{figure}
\vskip50pt
\centerline{\psfig{figure=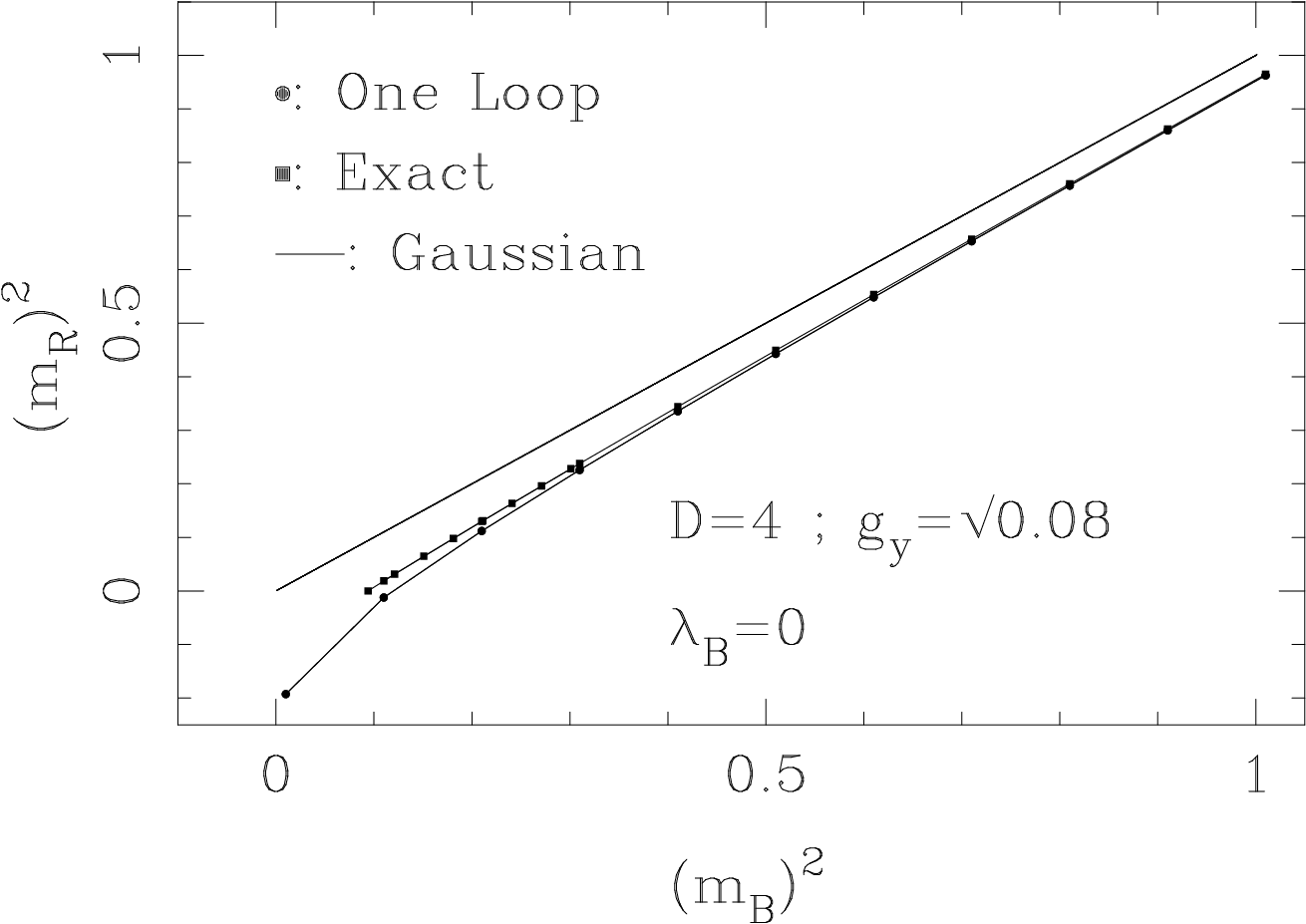,height=5in,angle=270}}
\vskip50pt
\caption{
The renormalized mass as a function of the bare mass in a bosonic $O(2)$
model with bare quartic coupling fixed to 0 and a Yukawa coupling equal to
$\sqrt{0.08}$}.
\vskip50pt
\end{figure}
\vfill
\eject
\begin{figure}
\vskip50pt
\centerline{\psfig{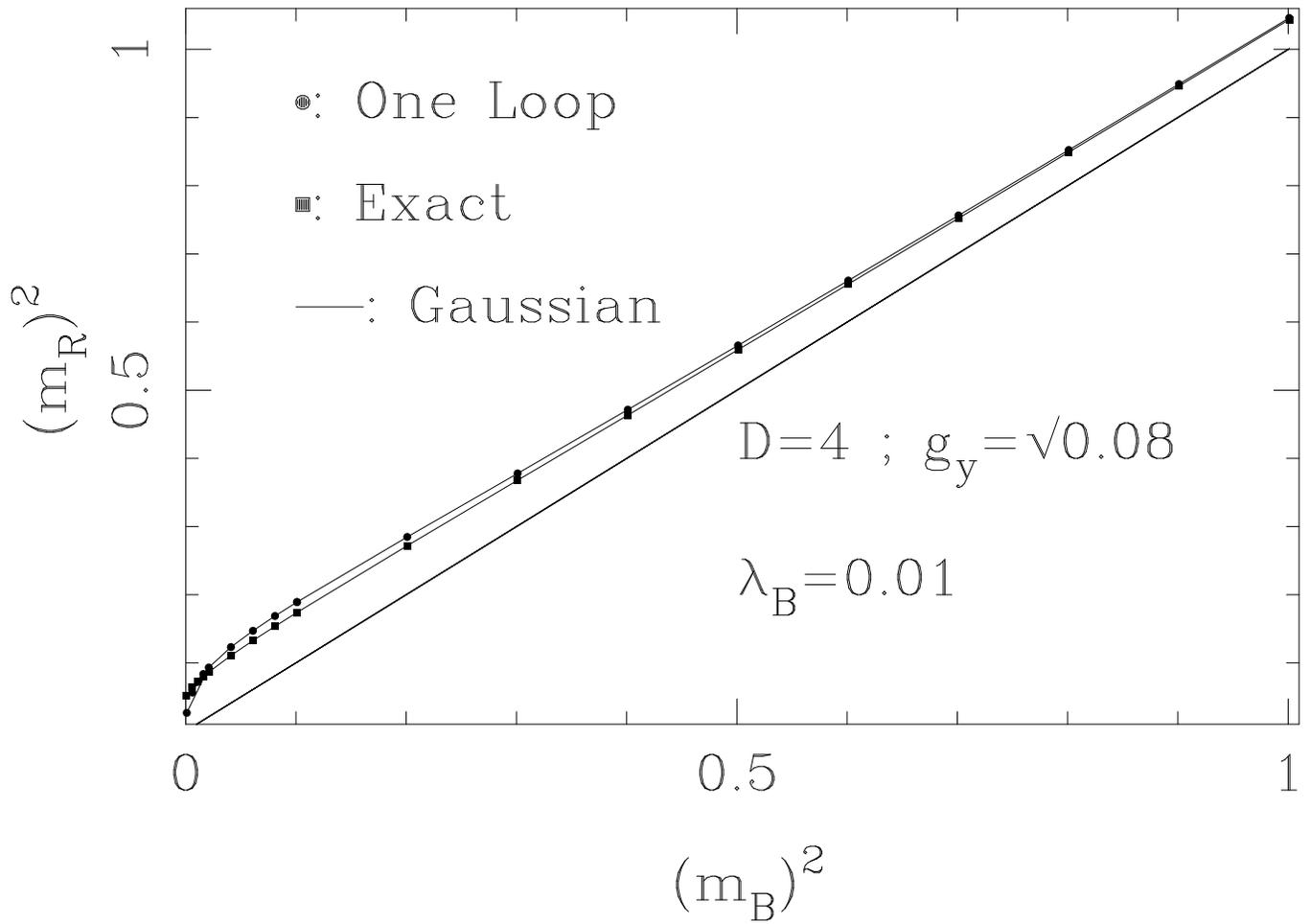}}
\vskip50pt
\caption{
The renormalized mass as a function of the bare mass in a bosonic $O(2)$
model with bare quartic coupling fixed to 0.01 and a Yukawa coupling equal to
$\sqrt{0.08}$.
}
\end{figure}
\vfill
\eject
\begin{figure}
\vskip50pt
\centerline{\psfig{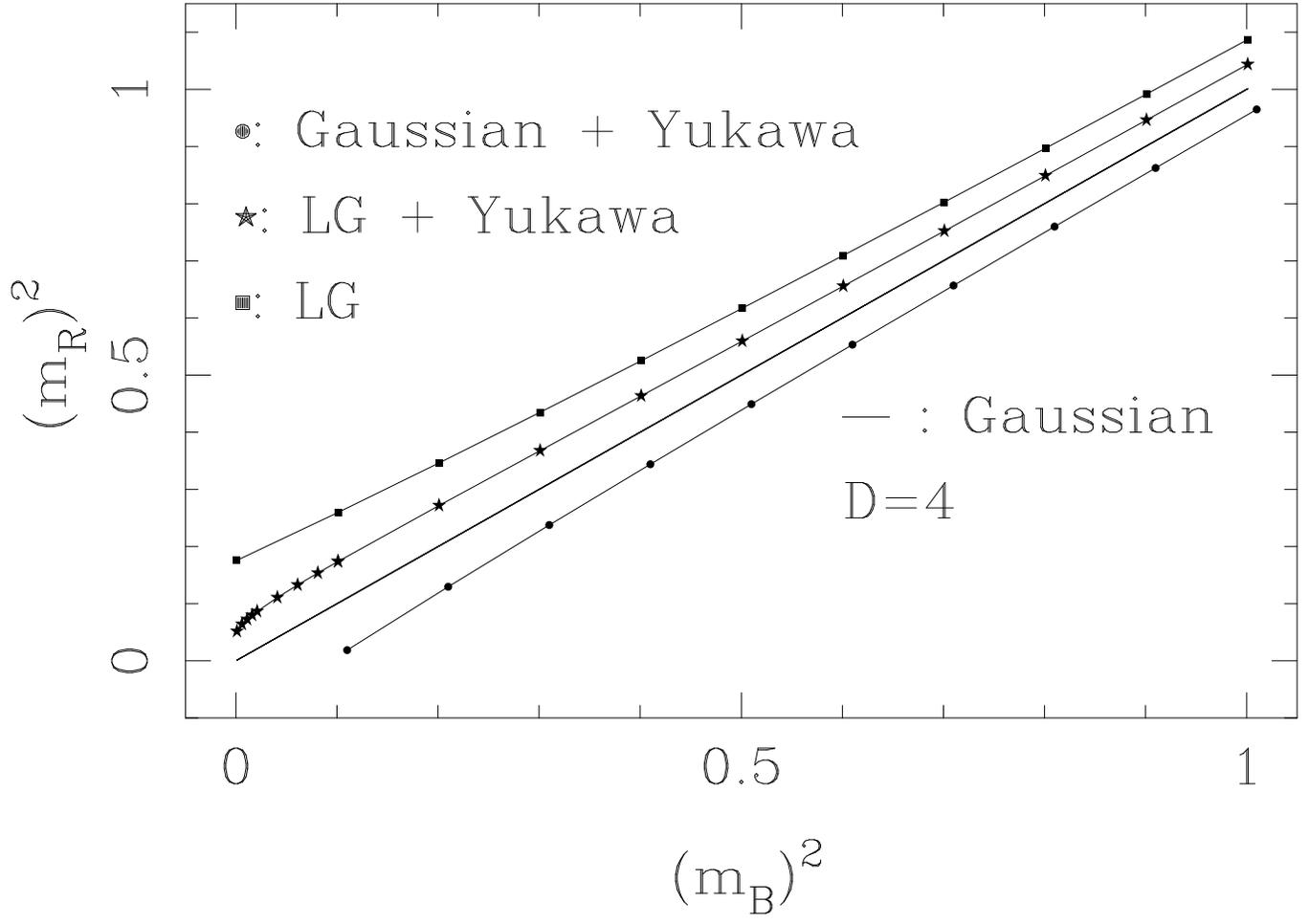}}
\vskip50pt
\caption{
The numerical results of Figs. 1, 2 and 3 combined together.
}
\end{figure}
\vfill
\eject
\begin{figure}\vskip50pt

\centerline{\psfig{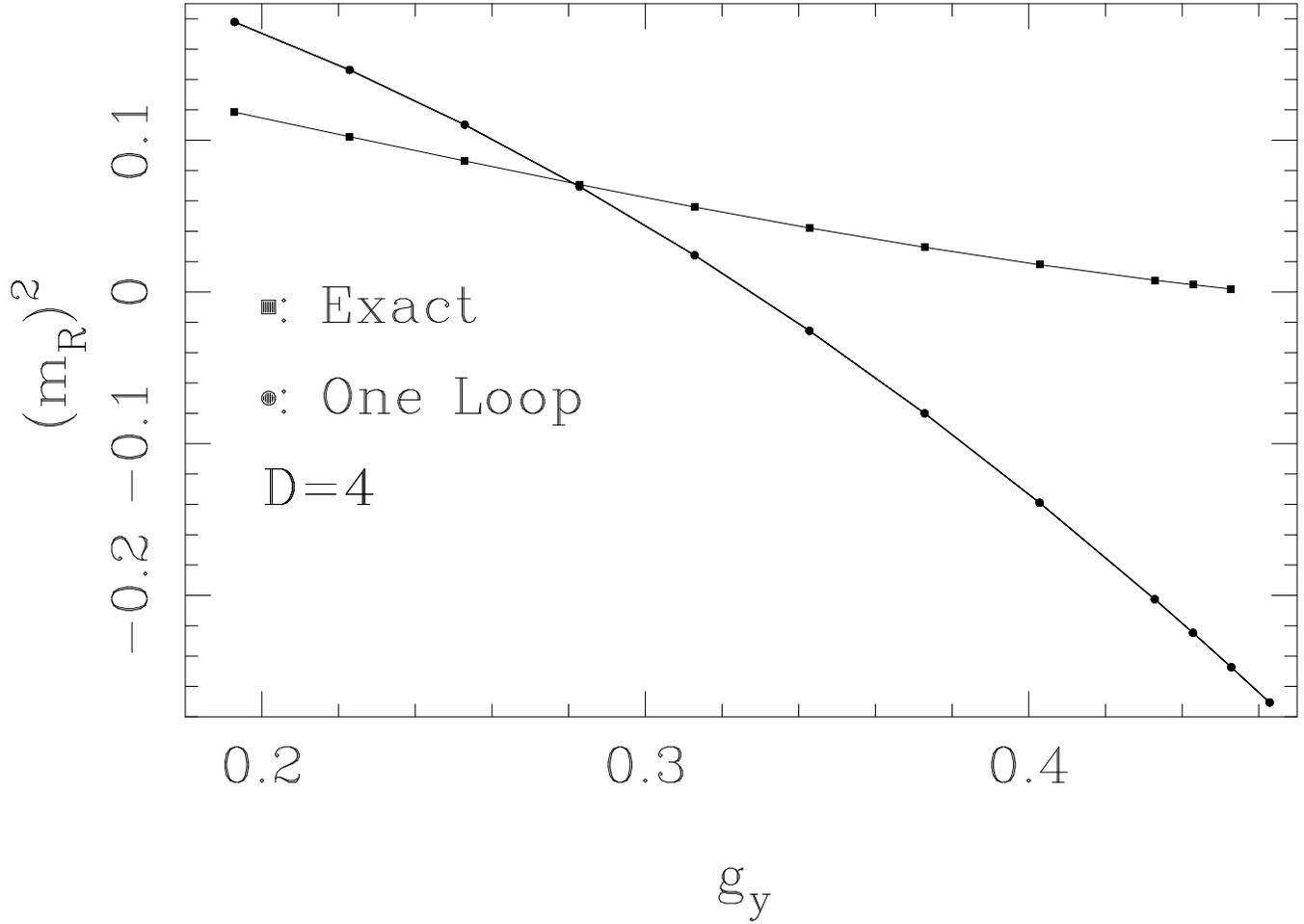}}
\vskip50pt
\caption{
The renormalized mass as a function of the Yukawa coupling in a bosonic $O(2)$
model with bare quartic coupling fixed to 0.01 and a bare mass $m_B^2=0.01$.
}
\end{figure}


\begin{references}
\bibitem{susskind}
L. Susskind,  Phys.\ Rev.\ D. {\bf 20 } 2619 (1979).
\bibitem{wilson}
K. Wilson, Phys.\ Rev.\ B. {\bf 4}, 3185 (1971) ; 
Phys.\ Rev.\ D. {\bf 3}, 1818 (1971);
K. Wilson and J. Kogut Phys.\ Rep. \ {\bf 12}, 75 (1974);
K.Wilson, Phys.\ Rev.\ D  {\bf 6}, 419 (1972).
%
\bibitem{wz}
J. Wess and B. Zumino, Nucl.\ Phys.\ B, {\bf 70 } 39 (1974).
\bibitem{dyson}
F. Dyson, Comm.\ Math.\ Phys.\ {\bf 12}, 91 (1969) ; 
G. Baker, Phys.\ Rev.\ B{\bf 5}, 2622 (1972);
G. Baker and G. Golner, Phys.\ Rev.\ B {\bf 16}, 2081 (1977);
Kim and Thomson, J. Phys. A {\bf 10}, 1579 (1977);
P. ~Bleher and Y. ~Sinai, Comm.\ Math.\ Phys.\ {\bf45}, 
247 (1975) ; P.~Collet and
J. P. ~Eckmann, Comm.\ Math.\ Phys.\ {\bf55}, 67 (1977);
H. Koch and P. Wittwer, Comm.\ Math.\ Phys.\ {\bf106}, 495 (1986) , 
{\bf 138}  537  (1991), 
{\bf 164}  627 (1994);
Y. Meurice, G. Ordaz and V.\ G.\ J.\ Rodgers, Phys.\ Rev.\ Lett. {\bf 75}, 
4555 (1995) ;
Y. Meurice, S. Niermann, and G. Ordaz, J.\ Stat.\ Phys.\ {\bf 87}, 363 (1997).

\bibitem{finite}
J.J Godina, Y. Meurice, B. Oktay and S. Niermann,
Phys.\ Rev.\ D  {\bf 57}, 6326 (1998).
\bibitem{jmp}
Y. Meurice, Jour. Math. Phys. {\bf 36}, 1812 (1995).
\bibitem{long}
J.J Godina, Y. Meurice, B. Oktay and S. Niermann,
Phys.\ Rev.\ D  {\bf 57}, R6581 (1998) and 
Phys.\ Rev.\ D  {\bf 59}, 096002 (1999).
\end{references}
\end{document}